\documentclass[10pt,final,doublecolumn]{IEEEtran}
\usepackage{amsmath}
\usepackage{latexsym}
\usepackage{graphicx}
\usepackage{bbding}
\usepackage{enumerate}
\usepackage{multicol}
\usepackage{color}
\usepackage{slashbox}
\IEEEoverridecommandlockouts
\begin{document}

\title{Performance Analysis and Optimization for Interference Alignment over MIMO Interference Channels with Limited Feedback}
\author{\normalsize
Xiaoming~Chen, \emph{Member, IEEE}, and Chau~Yuen, \emph{Senior
Member, IEEE}
\thanks{Copyright (c) 2013 IEEE. Personal use of this material
is permitted. However, permission to use this material for any
other purposes must be obtained from the IEEE by sending a request
to pubs-permissions@ieee.org.}
\thanks{This work was supported by the Natural
Science Funding of China (No. 61301102), Natural Science Funding
of Jiangsu Province (No. BK20130820), the open research fund of
National Mobile Communications Research Laboratory, Southeast
University (No. 2012D16), the Doctoral Fund of Ministry of
Education of China (No. 20123218120022) and Singapore University
of Technology and Design.}
\thanks{Xiaoming~Chen ({\tt chenxiaoming@nuaa.edu.cn}) is with the
College of Electronic and Information Engineering, Nanjing
University of Aeronautics and Astronautics, and also with the
National Mobile Communications Research Laboratory, Southeast
University, China. Chau~Yuen ({\tt yuenchau@sutd.edu.sg}) is with
the Singapore University of Technology and Design,
Singapore.}}\maketitle

\begin{abstract}
In this paper, we address the problem of interference alignment (IA)
over MIMO interference channels with limited channel state
information (CSI) feedback based on quantization codebooks. Due to
limited feedback and hence imperfect IA, there are residual
interferences across different links and different data streams. As
a result, the performance of IA is greatly related to the CSI
accuracy (namely number of feedback bits) and the number of data
streams (namely transmission mode). In order to improve the
performance of IA, it makes sense to optimize the system parameters
according to the channel conditions. Motivated by this, we first
give a quantitative performance analysis for IA under limited
feedback, and derive a closed-form expression for the average
transmission rate in terms of feedback bits and transmission mode.
By maximizing the average transmission rate, we obtain an adaptive
feedback allocation scheme, as well as a dynamic mode selection
scheme. Furthermore, through asymptotic analysis, we obtain several
clear insights on the system performance, and provide some
guidelines on the system design. Finally, simulation results
validate our theoretical claims, and show that obvious performance
gain can be obtained by adjusting feedback bits dynamically or
selecting transmission mode adaptively.
\end{abstract}


\section{Introduction}
The pioneer works by Cadambe \cite{IA1} and Maddah-Ali \cite{IA2}
spur considerable researches on interference alignment (IA), which
can effectively mitigate the interference over MIMO interference
channel and thus improve the performance \cite{IA3}-\cite{IA5}.
The principle of IA is to align the interferences from different
sources in some specific directions, so that the desired signal
can be transmitted without interference in a larger space. With
respect to other interference mitigation techniques, such as
zero-forcing beamforming (ZFBF) \cite{ZFBF}, IA increases the
spatial degrees of freedom, so it can accommodate more
transmission links, especially in the high signal-to-noise ratio
(SNR) region.

Previous analogous works mainly focus on the asymptotic performance
analysis and algorithm design of IA over MIMO interference channels
by assuming that infinity-approaching SNR. Since the capacity of
interference channel is still an open problem \cite{Capacity1}
\cite{Capacity2}, most works turn to the analysis of multiplexing
gain. It has been proved from the information-theoretic perspective
that IA can achieve at most $KM/2$ degrees of freedom (DOF) over
MIMO interference channels with $K$ transmitter-receiver links each
employing $M$ antennas \cite{IA1}. However, for general MIMO
interference channels, IA algorithm is unavailable when $K>3$,
except the numerical approach \cite{NumericalApp}. Only in some
special cases, IA algorithms that approach to interference-free DOF
are found. For example, a subspace interference alignment scheme
suitable for uplink cellular networks is proposed in \cite{Uplink}.
Then, the authors also present a bi-precoder IA scheme for downlink
cellular networks, which provides four-fold gain in throughput
performance over a standard multiuser MIMO technique
\cite{Downlink}.

A common disadvantage of the above IA schemes lies in that global
channel state information (CSI) must be available at each
transmitter, which weakens their applications in practical systems,
because CSI, especially interference CSI, is difficult to obtain at
the transmitter. In order to solve this challenging problem, the
authors in \cite{ChannelReciprocity} propose to perform
opportunistic IA by making use of channel reciprocity, but it is
only applicable to time division duplex (TDD) systems. In
\cite{PartialCSI}, a lattice interference alignment is proposed,
which only requires partial CSI. Moreover, blind IA without any CSI
is realized in some special cases \cite{BlindIA}. However, there is
obvious performance loss with respect to IA with full CSI.

In traditional MIMO systems, limited feedback based on quantization
codebook is a common and powerful method to aid the transmitters to
obtain the CSI from the receivers \cite{Overview}. Similarly, for IA
over MIMO interference channels, limited feedback scheme is also
viable \cite{LimitedFeedback1} \cite{LimitedFeedback0}. In
\cite{LimitedFeedback2}, Grassmannian manifold based limited
feedback technique is introduced into MIMO interference channels,
and the relationship between the performance of IA and the feedback
amount or codebook size is revealed. It is found that even with
limited CSI feedback, the full sum degrees of freedom of the
interference channel can be achieved. The authors in
\cite{LimitedFeedback3} make use of limited feedback theory to
analyze the performance of subspace IA in uplink cellular systems.
Furthermore, the subspace IA scheme with limited feedback is
optimized by minimizing the chordal distance of real CSI and
Grassmannian quantization codeword in \cite{LimitedFeedback4}; and
the outage capacity is analyzed for MIMO interference channel
employing IA with limited feedback in \cite{LimitedFeedback5}.

For IA based on limited CSI feedback, the residual interference (due
to imperfect IA) results in performance degradation with respect to
the case with perfect CSI \cite{LimitedFeedback2}. In order to
minimize the performance loss, it is necessary to take some
effective performance optimization measures. An upper bound on rate
loss caused by limited feedback is derived, and a beamformer design
method is given to minimize the upper bound in \cite{RateLoss}. In a
MIMO interference network, the residual interferences from different
transmitters are independent to each another, and have different
impacts on the performance. Considering that the total feedback
amount is constrained in practical system (due to limited feedback
capacity), in order to improve the system performance, we should
distribute the feedback resource among the forward and interference
channels according to channel conditions. For example, the authors
in \cite{FeedbackAllocation1} present a feedback allocation scheme
for IA in limited feedback MIMO interference channel with single
data stream for each link by minimizing the average residual
interference. Then, the feedback allocation scheme is extended to
the case with multiple data streams \cite{FeedbackAllocation2}.

Since the average residual interference is not directly related to
performance metric (e.g. transmission rate), feedback allocation
based on the criterion of minimizing the average residual
interference may be suboptimal. As widely known, the capacity of
interference channel is still an open issue, especially in the case
of limited feedback, so it is a challenging task to perform feedback
allocation from the perspective of maximizing the average
transmission rate directly. Moreover, the number of data streams,
namely transmission mode, also has a great impact on the
transmission rate together with feedback bits, especially in the
case of limited feedback. Specifically, a large number of data
streams can exploit more multiplexing gain, but also results in
higher residual interference. In fact, it has been proved that
dynamic mode selection is an effective way of improving the
performance for some MIMO systems, e.g. in multiuser MIMO systems,
several mode selection schemes have been proposed to optimize the
overall performance \cite{MS} \cite{Modeselection}.

Motivated by the above observations, we look into the matter of
performance analysis and optimization for IA with limited feedback
over a general MIMO interference channel. We assume each
transmitter-receiver MIMO channel can have a different path loss,
and has distinct number of data streams. The focus of this paper is
on analyzing the average transmission rate in terms of feedback
amount and transmission mode for IA over MIMO interference channels,
and then derives the corresponding adaptive feedback allocation and
dynamic mode selection schemes to optimize the performance. The
major contributions of this paper are summarized as follows:
\begin{enumerate}

\item We build a performance analysis framework for IA with
limited CSI feedback over MIMO interference channels, and derive a
closed-form expression for the average transmission rate in terms of
feedback amount and transmission mode.

\item We design an adaptive feedback allocation scheme by maximizing
the average transmission rate. Simulation results show that it poses
obvious performance gain over the baseline schemes.

\item We propose a dynamic mode selection scheme, namely choosing
the optimal number of data streams for each transmitter-receiver
link, so as to further optimize the performance.

\item We perform asymptotic analysis on the average transmission
rate, and obtain several insights, which can be served as guidelines
on the system design as follows:

\begin{enumerate}
\item Limited CSI feedback results in rate loss, and a performance
ceiling is created. The rate loss is an increasing function of
transmit power and a decreasing function of feedback amount. In
order to keep a constant gap with respect to IA with full CSI,
feedback amount should be increased as transmit power grows.

\item The larger the antenna number, the lower the CSI accuracy.
Hence, a large number of antennas may not lead to performance
improvement, if the feedback amount is not increased with the number
of antennas.

\item In interference-limited scenarios, single data stream for each
transmit-receive pair is optimal. While in noise-limited cases,
maximum feasible number of data streams should be chosen.

\item Under the noise-limited condition, CSI is useless for performance
improvement. In other word, CSI feedback is not necessary.

\end{enumerate}

\end{enumerate}

The rest of this paper is organized as follows: Section II gives a
brief introduction of the considered MIMO interference network with
limited feedback and IA. Section III focuses on performance analysis
of IA, and proposes a feedback allocation scheme as well as a mode
selection scheme. Section IV derives the average transmission rates
in two extreme cases through asymptotic analysis, and presents some
system design guidelines. Section V provides simulation results to
validate the effectiveness of the proposed schemes. Finally, Section
VI concludes the whole paper.

\emph{Notations}: We use bold upper (lower) letters to denote
matrices (column vectors), $(\cdot)^H$ to denote conjugate
transpose, $E[\cdot]$ to denote expectation, $\|\cdot\|$ to denote
the  $L_2$-norm of a vector, $|\cdot|$ to denote the absolute value,
$(a)^{+}$ to denote $\max(a,0)$, $\lceil a\rceil$ to denote the
smallest integer not less than $a$, $\lfloor a\rfloor$ to denote the
largest integer not greater than $a$, $\textmd{vec}(\textbf{A})$ to
denote matrix vectorization, $\stackrel{d}{=}$ to denote the
equality in distribution, and $\mathcal{O}(x)$ to denote increasing
proportionally with $x$. The acronym i.i.d. means ``independent and
identically distributed", pdf means ``probability density function"
and cdf means ``cumulative distribution function".

\section{System Model}
\begin{figure}[h] \centering
\includegraphics [width=0.45\textwidth] {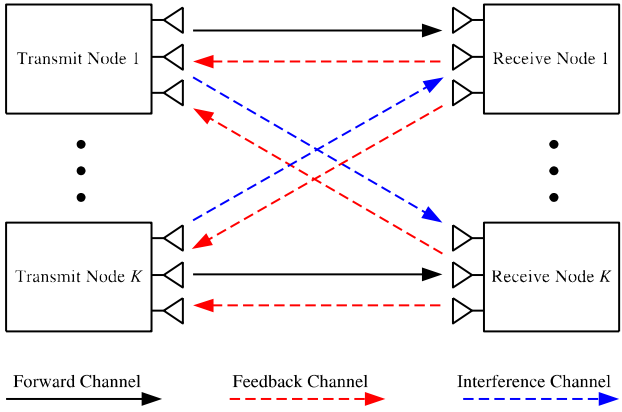}
\caption {A model of MIMO interference network with limited CSI
feedback.} \label{Fig1}
\end{figure}

We consider a MIMO interference network with $K$
transmitter-receiver links, as shown in Fig.\ref{Fig1}. For
convenience of analysis, we assume a homogeneous system, where all
transmitters and receivers are equipped with $N_t$ and $N_r$
antennas, respectively. While transmitter $k$ sends the signal to
its intended receiver $k$, it also creates interference to other
$K-1$ unintended receivers. We use
$\alpha_{k,i}^{1/2}\textbf{H}_{k,i}$ to denote the channel from
transmitter $i$ to receiver $k$, where $\alpha_{k,i}$ represents the
path loss and $\textbf{H}_{k,i}$ is the $N_r\times N_t$ fast fading
channel matrix with independent and identically distributed (i.i.d.)
zero mean and unit variance complex Gaussian entries. Transmitter
$k$ has $d_k$ independent data streams to be transmitted. It is
worth pointing out that due to the limitation of spatial degree of
freedom, the values of $d_k$s must fulfill the feasibility
conditions of IA \cite{Feasibility1} \cite{Feasibility2}. In what
follows, we assume IA is feasible by choosing $d_k$s carefully.
Thus, the received signal at receiver $k$ can be expressed as
\begin{eqnarray}
\textbf{y}_k
&=&\sum\limits_{i=1}^{K}\sqrt{\frac{P\alpha_{k,i}}{d_i}}\textbf{H}_{k,i}\sum\limits_{l=1}^{d_i}\textbf{w}_{i,l}s_{i,l}+\textbf{n}_{k},\label{eqn1}
\end{eqnarray}
where $\textbf{y}_k$ is the $N_r$ dimensional received signal
vector, $\textbf{n}_k$ is the additive Gaussian white noise with
zero mean and covariance matrix $\sigma^2\textbf{I}_{N_r}$,
$s_{i,l}$ denotes the $l$th normalized data stream from transmitter
$i$, and $\textbf{w}_{i,l}$ is the corresponding $N_t$ dimensional
beamforming vector. $P$ is the total transmit power at each
transmitter, which is equally allocated to its data streams.
Receiver $k$ uses the received vector $\textbf{v}_{k,j}$ of unit
norm to detect its $j$th data stream, which is given by
\begin{eqnarray}
\hat{s}_{k,j}&=&\textbf{v}_{k,j}^{H}\textbf{y}_k\nonumber\\
&=&\sqrt{\frac{P\alpha_{k,k}}{d_k}}\textbf{v}_{k,j}^{H}\textbf{H}_{k,k}\textbf{w}_{k,j}s_{k,j}\nonumber\\
&&+\sqrt{\frac{P\alpha_{k,k}}{d_k}}\sum\limits_{l=1,l\neq
j}^{d_k}\textbf{v}_{k,j}^{H}\textbf{H}_{k,k}\textbf{w}_{k,l}s_{k,l}\nonumber\\
&&+\sum\limits_{i=1,i\neq
k}^{K}\sqrt{\frac{P\alpha_{k,i}}{d_i}}\sum\limits_{l=1}^{d_i}\textbf{v}_{k,j}^{H}\textbf{H}_{k,i}\textbf{w}_{i,l}s_{i,l}+\textbf{v}_{k,j}^{H}\textbf{n}_{k},\label{eqn2}
\end{eqnarray}
where the first term at the right side of (\ref{eqn2}) is the
desired signal, the second one is the inter-stream interference
caused by the same transmitter, and the third one is the inter-link
interference resulting from the other transmitters. In order to
mitigate these interferences and improve the performance, IA is
performed accordingly. If perfect CSI is available at all nodes, we
have
\begin{equation}
\textbf{v}_{k,j}^{H}\textbf{H}_{k,k}\textbf{w}_{k,l}=0,\quad l\neq
j, \forall k\in[1,K], \forall l,j\in[1,d_k].\label{eqn3}
\end{equation}
and
\begin{equation}
\textbf{v}_{k,j}^{H}\textbf{H}_{k,i}\textbf{w}_{i,l}=0,\quad i\neq
k, \forall k,i\in[1,K], \forall l\in[1,d_i], \forall
j\in[1,d_k].\label{eqn4}
\end{equation}
In brief, inter-stream and inter-link interferences can be canceled
completely if perfect CSI is available. However, in practical
systems (e.g. frequency division duplex system), it is difficult for
the transmitters to obtain full CSI, including the interference CSI.
Since the feedback channel has limited bandwidth, codebook based
quantization is an effective way to convey partial CSI from the
receivers to the transmitters. In a MIMO interference network, CSI
is quantized in the form of vectorization. Specifically, for the
channel $\textbf{H}_{k,i}$, it is first vectorized as
$\textbf{h}_{k,i}=\textmd{vec}(\textbf{H}_{k,i})$, then receiver $k$
selects an optimal codeword from a predetermined codebook
$\mathcal{H}_{k,i}=\left\{\hat{\textbf{h}}_{k,i}^{(1)},\cdots,\hat{\textbf{h}}_{k,i}^{(2^{B_{k,i}})}\right\}$
of size $2^{B_{k,i}}$ according to the following criterion:
\begin{equation}
b^{\star}=\arg\max\limits_{1\leq
b\leq2^{B_{k,i}}}\left|\tilde{\textbf{h}}_{k,i}^{H}\hat{\textbf{h}}_{k,i}^{(b)}\right|^2,\label{eqn5}
\end{equation}
where
$\tilde{\textbf{h}}_{k,i}=\textbf{h}_{k,i}/{\left\|\textbf{h}_{k,i}\right\|}$
is the channel direction vector. Transmitter $i$ recovers the
quantized CSI from the same codebook $\mathcal{H}_{k,i}$ after
receiving the feedback information $b^{\star}$, and then constructs
its beamforming vectors $\textbf{w}_{i,l},l=1,\cdots, d_i$ based on
all the feedback information about the related forward and
interference channels, namely $\textbf{H}_{i,k}, \forall
k=1,\cdots,K$.

The MIMO interference network is operated in slotted time. At the
beginning of each time slot, receiver $k$ conveys the optimal
codeword index to transmitter $i$ with $B_{k,i}$ bits, where
$i=1,\cdots,K$. It is worth pointing out that we consider a low
mobility scenario, so the impact of feedback delay is negligible.
Due to the limitation of feedback resource, we assume each receiver
has $B$ feedback bits in total during one time slot. The focus of
this paper is on performance analysis and optimization subject to
$\sum\limits_{i=1}^KB_{k,i}=B$ for $k=1,\cdots,K$.

\section{Performance Analysis and Optimization}
In this section, we concentrate on performance analysis and
optimization for IA in a MIMO interference network with limited CSI
feedback. In the case of limited CSI feedback, the capacity of MIMO
interference network is still an open issue. For technical
tractability, we alternatively put the attention on the average
transmission rate. In the sequence, we first give a detailed
investigation of average transmission rate, and then propose an
adaptive feedback allocation scheme as well as a dynamic mode
selection scheme for IA in a general MIMO interference network to
optimize the overall performance.

\subsection{Average Transmission Rate}
Due to limited CSI feedback, although IA is adopted, (\ref{eqn3})
and (\ref{eqn4}) do not hold true any more. These result in residual
interference, also called interference leakage. Under such
condition, the signal to interference plus noise ratio (SINR)
related to the $j$th data stream of transmitter-receiver link $k$
can be expressed as
\begin{eqnarray}
\gamma_{k,j}&=&\frac{\kappa_{k,k}\left|\textbf{v}_{k,j}^{H}\textbf{H}_{k,k}\textbf{w}_{k,j}\right|^2}{I_{k,j}+\sigma^2}\nonumber\\
&=&\frac{\kappa_{k,k}\left|\textbf{h}_{k,k}^H\textbf{T}_{j,j}^{(k,k)}\right|^2}{I_{k,j}+\sigma^2},\label{eqn24}
\end{eqnarray}
where $\kappa_{k,k}=\frac{P\alpha_{k,k}}{d_k}$,
$\textbf{T}_{j,j}^{(k,k)}=\textbf{w}_{k,i}\otimes\textbf{v}_{k,j}^H$
and $\otimes$ represents the Kronecker product. $I_{k,j}$ is the
total residual interference, which is given by
\begin{eqnarray}
I_{k,j}&=&\kappa_{k,k}\sum\limits_{l=1,l\neq
j}^{d_k}\left|\textbf{v}_{k,j}^{H}\textbf{H}_{k,k}\textbf{w}_{k,l}\right|^2\nonumber\\
&&+\sum\limits_{i=1,i\neq
k}^{K}\kappa_{k,i}\sum\limits_{l=1}^{d_i}\left|\textbf{v}_{k,j}^{H}\textbf{H}_{k,i}\textbf{w}_{i,l}\right|^2\nonumber\\
&=&\kappa_{k,k}\sum\limits_{l=1,l\neq
j}^{d_k}\left|\textbf{h}_{k,k}^H\textbf{T}_{j,l}^{(k,k)}\right|^2\nonumber\\
&&+\sum\limits_{i=1,i\neq
k}^{K}\kappa_{k,i}\sum\limits_{l=1}^{d_i}\left|\textbf{h}_{k,i}^H\textbf{T}_{j,l}^{(k,i)}\right|^2\nonumber\\
&=&\kappa_{k,k}\|\textbf{h}_{k,k}\|^2\sum\limits_{l=1,l\neq
j}^{d_k}\left|\tilde{\textbf{h}}_{k,k}^H\textbf{T}_{j,l}^{(k,k)}\right|^2\nonumber\\
&&+\sum\limits_{i=1,i\neq
k}^{K}\kappa_{k,i}\|\textbf{h}_{k,i}\|^2\sum\limits_{l=1}^{d_i}\left|\tilde{\textbf{h}}_{k,i}^H\textbf{T}_{j,l}^{(k,i)}\right|^2,\label{eqn6}
\end{eqnarray}
where $\kappa_{k,j}=\frac{P\alpha_{k,j}}{d_j}$ and
$\textbf{T}_{j,l}^{(k,i)}=\textbf{w}_{i,l}\otimes\textbf{v}_{k,j}^H$.
Following the theory of random vector quantization \cite{RVQ}, the
relation between the original channel direction vector
$\tilde{\textbf{h}}_{k,i}$, and the quantized channel direction
vector $\hat{\textbf{h}}_{k,i}$ can be expressed as
\begin{equation}
\tilde{\textbf{h}}_{k,i}=\sqrt{1-a_{k,i}}\hat{\textbf{h}}_{k,i}+\sqrt{a_{k,i}}\textbf{s}_{k,i},\label{eqn7}
\end{equation}
where
$a_{k,i}=\sin^2\left(\angle\left(\tilde{\textbf{h}}_{k,i},\hat{\textbf{h}}_{k,i}\right)\right)$
is the magnitude of the quantization error, and $\textbf{s}_{k,i}$
is an unit norm vector isotropically distributed in the nullspace of
$\hat{\textbf{h}}_{k,i}$, and is independent of $a_{k,i}$. Since IA
is performed based on the quantized CSI $\hat{\textbf{h}}_{k,i}$, so
we have
\begin{eqnarray}
\left|\tilde{\textbf{h}}_{k,i}^H\textbf{T}_{j,l}^{(k,i)}\right|^2&=&\left|\sqrt{1-a_{k,i}}\hat{\textbf{h}}_{k,i}^H\textbf{T}_{j,l}^{(k,i)}+\sqrt{a_{k,i}}\textbf{s}_{k,i}^H\textbf{T}_{j,l}^{(k,i)}\right|^2\nonumber\\
&=&a_{k,i}\left|\textbf{s}_{k,i}^H\textbf{T}_{j,l}^{(k,i)}\right|^2,\label{eqn8}
\end{eqnarray}
where (\ref{eqn8}) holds true according to the IA principles
(\ref{eqn3}) and (\ref{eqn4}). In this case, the residual
interference in (\ref{eqn6}) is reduced as
\begin{eqnarray}
I_{k,j}&=&\kappa_{k,k}a_{k,k}\|\textbf{h}_{k,k}\|^2\sum\limits_{l=1,l\neq
j}^{d_k}\left|\textbf{s}_{k,k}^H\textbf{T}_{j,l}^{(k,k)}\right|^2\nonumber\\
&&+\sum\limits_{i=1,i\neq
k}^{K}\kappa_{k,i}a_{k,i}\|\textbf{h}_{k,i}\|^2\sum\limits_{l=1}^{d_i}\left|\textbf{s}_{k,i}^H\textbf{T}_{j,l}^{(k,i)}\right|^2,\label{eqn9}
\end{eqnarray}
where the first term at the right side is the residual inter-stream
interference and the second term is the residual inter-link terms.
In fact, the two kinds of interferences are equivalent if we
consider each data stream as an independent link.

Hence, the average transmission rate for the $j$th data stream of
transmitter-receiver link $k$ can be computed as
\begin{eqnarray}
\bar{R}_{k,j}&&\!\!\!\!\!\!\!\!\!\!\!\!\!\!\!(\textbf{B}_k,\textbf{d})\nonumber\\&=&E\left[\log_2\left(1+\gamma_{k,j}\right)\right]\label{eqn54}\\
&=&E\left[\log_2\left(\kappa_{k,k}\left|\textbf{h}_{k,k}^H\textbf{T}_{j,j}^{(k,k)}\right|^2+I_{k,j}+\sigma^2\right)\right]\nonumber\\
&&-E\left[\log_2\left(I_{k,j}+\sigma^2\right)\right]\label{eqn25}\\
&=&\frac{1}{\ln2}\sum\limits_{i=1}^{L}\sum\limits_{t=1}^{\eta_{k,i}}\Xi_L\Bigg(i,t,\{\eta_{k,q}\}_{q=1}^{L},\nonumber\\
&&\left\{\frac{\varrho_{k,q}}{\eta_{k,i}}\right\}_{q=1}^{L},\{l_{k,q}\}_{q=1}^{L-2}\Bigg)Z\left(\sigma^2,t,\frac{\varrho_{k,i}}{\eta_{k,i}}\right)\nonumber\\
&&-\frac{1}{\ln2}\sum\limits_{i=1}^{K}\sum\limits_{t=1}^{\omega_{k,i}}\Xi_K\Bigg(i,t,\{\omega_{k,q}\}_{q=1}^{K},\nonumber\\
&&\left\{\frac{\varrho_{k,q}}{\omega_{k,i}}\right\}_{q=1}^{K},\{l_{k,q}\}_{q=1}^{K-2}\Bigg)Z\left(\sigma^2,t,\frac{\varrho_{k,i}}{\omega_{k,i}}\right),\label{eqn34}
\end{eqnarray}
where
\begin{eqnarray}
Z(x,y,z)&=&\ln(x)+\sum\limits_{\vartheta=0}^{y-1}\frac{1}{\Gamma(t-\vartheta)}\bigg((-1)^{y-\vartheta-2}\left(\frac{x}{z}\right)^{y-\vartheta-1}\nonumber\\
&&\times\exp\left(\frac{x}{z}\right)
\textmd{E}_{\textmd{i}}\left(-\frac{x}{z}\right)\nonumber\\&&
+\sum\limits_{\nu=1}^{y-\vartheta-1}\Gamma(\nu)\left(-\frac{x}{z}\right)^{y-\vartheta-\nu-1}\bigg),\nonumber
\end{eqnarray}
$\textmd{E}_{\textmd{i}}(x)=\int_{-\infty}^{x}\frac{\exp(t)}{t}dt$
is the exponential integral function,
$\textbf{B}_{k}=\{B_{k,1},\cdots,B_{k,K}\}$ is a certain feedback
bits allocation result related to receiver $k$, and
$\textbf{d}=\{d_1,\cdots,d_K\}$ is a combination set on the numbers
of data streams fulfilling the feasibility conditions
\cite{Feasibility1} \cite{Feasibility2}. The proof of the above
expression is presented in Appendix I.

\emph{Remark}: It is found that (\ref{eqn34}) is independent of the
data stream index $j$, since the desired signal quality and the
residual interference are the same for all data streams of link $k$
in statistical sense. Therefore, the total rate of link $k$ is $d_k$
times $\bar{R}_{k,j}$, and thus the sum of average transmission rate
for the MIMO interference network employing IA with limited CSI
feedback is given by
\begin{eqnarray}
\bar{R}(\textbf{B}_k,\textbf{d})&=&\sum\limits_{k=1}^{K}d_k\bar{R}_{k,j}(\textbf{B}_k,\textbf{d}).\label{eqn35}
\end{eqnarray}

\subsection{Adaptive Feedback Allocation}
As seen in (\ref{eqn34}), given channel conditions and the number of
data streams, average transmission rate is a function of feedback
bits $\textbf{B}_{k}$. In order to maximize the average transmission
rate, it is necessary to distribute the feedback bits at each
receiver, which is equivalent to the following optimization problem
\begin{eqnarray}
J_1\!\!\!\!\!&:&\!\!\!\!\!
\max\limits_{\textbf{B}_k}\bar{R}_{k,j}(\textbf{B}_k,\textbf{d})\label{eqn11}\\
&s.t.&\sum\limits_{i=1}^{K}B_{k,i}=B\label{eqn12}
\end{eqnarray}
Evidently, $J_1$ is an integer programming problem and
$\bar{R}_{k,j}(\textbf{B}_{k},\textbf{d})$ is a complicated function
of $\textbf{B}_k$, so it is difficult to obtain a closed-form
expression for the optimal solution. Intuitively, the optimal
results can be achieved by using the numerical searching method, but
the complexity increases proportionally with $K^B$, which is
unbearable in practical systems with large $K$ and $B$. In order to
get a balance between the performance and the complexity, we propose
a greedy scheme to allocate the feedback resource bit by bit, and
each bit is allocated to the channel that having $\bar{R}_{k,j}$
increases the fastest. The whole greedy feedback allocation scheme
can be summarized as follows:

\rule{8.5cm}{1pt}
\begin{enumerate}
\item Initialization: Given $N_t$, $N_r$, $K$, $B$, $P$,
$\sigma^2$, $d_i$ and $\alpha_{k,i}$ for $i=1,\cdots,K$. Let
$B_{k,1}=\cdots=B_{k,K}=0$ and
$\bar{R}_{k,j}(\textbf{B}_{k},\textbf{d})$ be defined as
(\ref{eqn34}). \item Let $\textbf{Q}^{(i)}=\{Q_1,\cdots,Q_{K}\}$,
where $Q_{j}=B_{k,j}$ for $j\neq i$, and $Q_i=B_{k,i}+1$. Search
$i^{\star}=\max\limits_{1\leq i\leq
K}\left(\bar{R}_{k,j}(\textbf{Q}^{(i)},\textbf{d})-\bar{R}_{k,j}(\textbf{B}_{k},\textbf{d})\right)$,
then let $B_{k,i^{\star}}=B_{k,i^{\star}}+1$, $B=B-1$ and update
$\textbf{B}_{k}$. \item If $B>0$, then go to 2). Otherwise,
$\textbf{B}_{k}$ is the feedback bits allocation result.
\end{enumerate}

\rule{8.5cm}{1pt}

Note that the proposed scheme distributes each feedback bit by
comparing $K$ rate increments, so the computational complexity is
$\mathcal{O}(KB)$, which is simpler than the numerical search
method. For an arbitrary receiver, the above scheme can also be used
to obtain the feedback bits allocation result by substituting the
corresponding network parameters.

\subsection{Dynamic Mode Selection}
As seen in (\ref{eqn35}), the number of data streams $d_i$ has a
great impact on the average transmission rate as well. While a
larger $d_i$ leads to a higher multiplex gain, it also results in
high interference. Hence, it is beneficial to select the optimal
number of data stream, namely mode selection, from the perspective
of the overall network performance. Taking the maximization of the
sum of average transmission rate as the optimization objective, the
problem of mode selection can be described as
\begin{eqnarray}
J_2\!\!\!\!\!&:&\!\!\!\!\!
\max\limits_{\textbf{d}}\bar{R}(\textbf{B}_{k},\textbf{d})\label{eqn36}\\
&s.t.&d_i\quad\text{satisfies the feasibility condition}\quad
\forall i.\label{eqn37}
\end{eqnarray}

So far, the necessary and sufficient condition for the feasibility
of IA for a general MIMO interference network is still an open
problem. Since only the sufficient condition in the symmetric MIMO
interference network is obtained, we consider the links that have
the same number of data streams $d$, and change the constraint
condition (\ref{eqn37}) as $N_t+N_r-(K+1)d\geq0$ according to
\cite{Feasibility1} and \cite{Feasibility2}. Thus, $J_2$, as an
integer optimization problem, can be solved by the numerical
searching method, and the total searching times is
$\left\lfloor\frac{N_t+N_r}{K+1}\right\rfloor$, which is not so
large in practical MIMO interference network with a limited number
of antennas. For example, the maximum number of antennas for the
LTE-A system is 8. Even if the link number is 4, the total searching
number times is only 3. In particular, as verified by theoretical
analysis and numerical simulation in the rest of this paper, the
MIMO interference network either chooses $d=1$ or adopts the maximum
feasible mode
$d_{\max}=\left\lfloor\frac{N_t+N_r}{K+1}\right\rfloor$. Thus, we
only need to compare the two transmission modes with bearable
complexity.

\subsection{Joint Optimization Scheme}
Feedback allocation and mode selection are integrally related. To be
precise, given a feedback allocation result, there exists an optimal
transmission mode combination. Similarly, a transmission mode
combination corresponds to an optimal feedback allocation result.
Hence, it is imperative to jointly optimize the two schemes, so as
to maximize the sum of average transmission rate. In the sequence,
we give a joint optimization scheme based on iteration as follows

\rule{8.5cm}{1pt}
\begin{enumerate}
\item Initialization: Given $N_t$, $N_r$, $K$, $B$, $P$,
$\sigma^2$ and $\alpha_{k,i}$ for $i=1,\cdots,K$. Let
$d_1=\cdots=d_K=d=1$ and $B_{k,1}=\cdots=B_{k,K}=0$;

\item Given $\textbf{d}$, perform feedback allocation to obtain
$\textbf{B}_k, \forall k\in[1,K]$;

\item Given $\textbf{B}_k$s, perform mode selection
through searching $d$ from 1 to $d_{\max}$ to obtain $\textbf{d}$;

\item If nether $\textbf{B}_k$s nor $\textbf{d}$ converge, go to 2).
Otherwise, $\textbf{B}_k$s and $\textbf{d}$ are the joint
optimization results.
\end{enumerate}

\rule{8.5cm}{1pt}

\section{Asymptotic Analysis}
In the last section, we have successfully derived the closed-form
expression of the average transmission rate for IA with limited
feedback in MIMO interference network, and presented two performance
optimization schemes, namely adaptive feedback allocation and
dynamic mode selection. A potential drawback is the high complexity
of the expression and thus the optimization schemes. In order to
obtain some insights on the system performance and hence extract
several simple design guidelines, we carry out asymptotic
performance analysis in two extreme cases, i.e. interference limited
and noise limited. In what follows, we give a detailed investigation
of average transmission rate and the corresponding performance
optimization schemes in the two cases, respectively.

\subsection{Interference Limited Case}
If transmit power $P$ is large enough, the noise term of SINR in
(\ref{eqn24}) can be negligible, thus the average transmission rate
for the $j$th data stream of link $k$ is reduced as
\begin{eqnarray}
\bar{R}_{k,j}&&\!\!\!\!\!\!\!\!\!\!\!\!\!\!\!(\textbf{B}_k,\textbf{d})\nonumber\\&=&E\left[\log_2\left(\kappa_{k,k}\left|\textbf{h}_{k,k}^H\textbf{T}_{j,j}^{(k,k)}\right|^2+I_{k,j}\right)\right]\nonumber\\
&&-E\left[\log_2\left(I_{k,j}\right)\right]\label{eqn38}\\
&=&\frac{1}{\ln2}\bigg(\sum\limits_{i=1}^{L}\sum\limits_{t=1}^{\eta_{k,i}}\Xi_L\Bigg(i,t,\{\eta_{k,q}\}_{q=1}^{L},\left\{\frac{\varrho_{k,q}}{\eta_{k,i}}\right\}_{q=1}^{L-2},\nonumber\\
&&\{l_{k,q}\}_{q=1}^{L-2}\Bigg)\left(\psi(t)+\ln(\varrho_{k,i})-\ln(\eta_{k,i})\right)\nonumber
\end{eqnarray}
\begin{eqnarray}
&-&\sum\limits_{i=1}^{K}\sum\limits_{t=1}^{\omega_{k,i}}\Xi_K\Bigg(i,t,\{\omega_{k,q}\}_{q=1}^{K},\left\{\frac{\varrho_{k,q}}{\omega_{k,i}}\right\}_{q=1}^{K},\nonumber\\
&&\{l_{k,q}\}_{q=1}^{K-2}\Bigg)\left(\psi(t)+\ln(\varrho_{k,i})-\ln(\omega_{k,i})\right)\bigg),\label{eqn41}
\end{eqnarray}
where the proof of the above expression is presented in Appendix II.
Then, the average transmission rate for the whole network is given
by
\begin{eqnarray}
\bar{R}&&\!\!\!\!\!\!\!\!\!\!\!\!\!\!\!(\textbf{B}_k,\textbf{d})\nonumber\\&=&\sum\limits_{k=1}^{K}\frac{d_k}{\ln2}\bigg(\sum\limits_{i=1}^{L}\sum\limits_{t=1}^{\eta_{k,i}}\Xi_L\Bigg(i,t,\{\eta_{k,q}\}_{q=1}^{L},\nonumber\\
&&\left\{\frac{\varrho_{k,q}}{\eta_{k,i}}\right\}_{q=1}^{L-2}\Bigg)\left(\psi(t)+\ln(\varrho_{k,i})-\ln(\eta_{k,i})\right)\nonumber\\
&&-\sum\limits_{i=1}^{K}\sum\limits_{t=1}^{\omega_{k,i}}\Xi_K\Bigg(i,t,\{\omega_{k,q}\}_{q=1}^{K},\left\{\frac{\varrho_{k,q}}{\omega_{k,i}}\right\}_{q=1}^{K},\nonumber\\
&&\{l_{k,q}\}_{q=1}^{K-2}\Bigg)\left(\psi(t)+\ln(\varrho_{k,i})-\ln(\omega_{k,i})\right)\bigg).\label{eqn42}
\end{eqnarray}
Substituting (\ref{eqn41}) into (\ref{eqn11}), and (\ref{eqn42})
into (\ref{eqn36}), we can get the corresponding feedback allocation
and mode selection schemes.

Note that
$\ln(\varrho_{k,i})=\ln(P)+\ln\left(\alpha_{k,i}2^{-\frac{B_{k,i}}{N_tN_r-1}}/d_i\right)$
$\forall i\in[1,K]$,
$\ln(\varrho_{k,L})=\ln(P)+\ln\left(\alpha_{k,k}/d_k\right)$ and
$\sum\limits_{i=1}^{L}\sum\limits_{t=1}^{\eta_{k,i}}\Xi_L\left(i,t,\{\eta_{k,q}\}_{q=1}^{L},\left\{\frac{\varrho_{k,q}}{\eta_{k,i}}\right\}_{q=1}^{L-2}\right)=1
$ for arbitrary $L$ because of $\int_0^{\infty}f(x)=1$, so the terms
related to $\ln(P)$ in $H(\textbf{B}_k,\textbf{d})$ and
$G(\textbf{B}_k,\textbf{d})$ cancel out. Thus, we have the following
theorem:

\emph{Theorem 1}: In the region of high transmit power, the average
transmission rate is independent of $P$, and there exists a
performance ceiling regardless of $P$, i.e. once $P$ is larger than
a saturation point, the average transmission rate will not increase
further even the transmit power increases.

Intuitively, the performance ceiling is an increasing function of
feedback bit $B$. To be precise, with the increase of $B$, the
ceiling rises accordingly. Once $B$ is large enough, resulting in
high SINR, then the constant term 1 in (\ref{eqn54}) can be
negligible, then the average transmission rate can be approximated
as
\begin{eqnarray}
\bar{R}_{k,j}(\textbf{B}_k,\textbf{d})&=&E\left[\log_2\left(\kappa_{k,k}\left|\textbf{h}_{k,k}^H\textbf{T}_{j,j}^{(k,k)}\right|^2\right)\right]\nonumber\\
&&-E\left[\log_2\left(I_{k,j}\right)\right],\label{eqn55}
\end{eqnarray}
If $B$ approaches infinity, namely perfect CSI at the transmitters,
then the interference can be cancelled completely by IA. In this
case, the average transmission rate can be expressed as
\begin{eqnarray}
\bar{R}_{k,j}(\infty,\textbf{d})&=&E\left[\log_2\left(\kappa_{k,k}\left|\textbf{h}_{k,k}^H\textbf{T}_{j,j}^{(k,k)}\right|^2\right)\right].\label{eqn56}
\end{eqnarray}
Therefore, the performance loss caused by limited CSI feedback under
the condition of large $P$ and $B$ is given by
\begin{eqnarray}
\Delta\bar{R}_{k,j}&&\!\!\!\!\!\!\!\!\!\!\!\!\!\!\!(\textbf{B}_k,\textbf{d})\nonumber\\&=&\bar{R}_{k,j}(\infty,\textbf{d})-\bar{R}_{k,j}(\textbf{B}_k,\textbf{d})\nonumber\\
&=&E\left[\log_2\left(I_{k,j}\right)\right]\nonumber\\
&=&\frac{1}{\ln(2)}\sum\limits_{i=1}^{K}\sum\limits_{t=1}^{\omega_{k,i}}\Xi_K\Bigg(i,t,\{\omega_{k,q}\}_{q=1}^{K},\left\{\frac{\varrho_{k,q}}{\omega_{k,i}}\right\}_{q=1}^{K},\nonumber\\
&&\{l_{k,q}\}_{q=1}^{K-2}\Bigg)\left(\psi(t)+\ln(\varrho_{k,i})-\ln(\omega_{k,i})\right)\nonumber\\
&\approx&\frac{1}{\ln(2)}\sum\limits_{i=1}^{K}\sum\limits_{t=1}^{\omega_{k,i}}\Xi_K\Bigg(i,t,\{\omega_{k,q}\}_{q=1}^{K},\left\{\frac{\varrho_{k,q}}{\omega_{k,i}}\right\}_{q=1}^{K},\nonumber\\
&&\{l_{k,q}\}_{q=1}^{K-2}\Bigg)\ln(\varrho_{k,i})\label{eqn57}\\
&=&\frac{1}{\ln(2)}\sum\limits_{i=1}^{K}\sum\limits_{t=1}^{\omega_{k,i}}\Xi_K\Bigg(i,t,\{\omega_{k,q}\}_{q=1}^{K},\left\{\frac{\varrho_{k,q}}{\omega_{k,i}}\right\}_{q=1}^{K},\nonumber\\
&&\{l_{k,q}\}_{q=1}^{K-2}\Bigg)\ln\left(P2^{-\frac{B_i}{N_tN_r-1}}\alpha_{k,i}/d_i\right),\label{eqn58}
\end{eqnarray}
where (\ref{eqn57}) follows the fact that
$\psi(t)-\ln(\omega_{k,i})$ is negligible with respect to
$\ln(\varrho_{k,i})$ when $P$ is large enough. Given $B$, the
performance loss will enlarge as $P$ increases (due to performance
ceiling mentioned above). In order to avoid the performance ceiling,
$P2^{-\frac{B_{k,i}}{N_tN_r-1}}$ should remain constant. Hence, we
have the following theorem:

\emph{Theorem 2}: If the number of feedback bits satisfies
$B=K(N_tN_R-1)\log_2\left(\frac{P}{\theta}\right)$, the performance
gap between the full and limited CSI remains constant, where
$\theta$ is a constant.

\begin{proof}
As analyzed above, as long as
$P2^{-\frac{B_{k,i}}{N_tN_r-1}}=\theta$ holds true, the performance
loss is a constant. Equivalently, we have
$B_{k,i}=(N_tN_r-1)\log_2\left(\frac{P}{\theta}\right)$, and then
$B=\sum\limits_{i=1}^KB_{k,i}=K(N_tN_R-1)\log_2\left(\frac{P}{\theta}\right)$.
\end{proof}

From Theorem 2, we know that in order for the sum of average
transmission rate under limited feedback to keep up the rate as
perfect CSI feedback, $B$ needs to be increased as $P$ increases.
Also from Theorem 2, it is found that with the increase of antenna
number, in order to keep the constant performance gap, it is
necessary to add more feedback bits. This is because when the number
of antenna becomes large, the CSI quantization accuracy decreases
accordingly, and there will be more residual interference.
Furthermore, if we assume all links have the same number of data
streams $d$, then the total performance loss due to limited CSI
feedback can be expressed as
\begin{eqnarray}
\Delta\bar{R}(\textbf{B}_k,\textbf{d})&=&d(\zeta_1-\zeta_2\ln(d)),\label{eqn59}
\end{eqnarray}
where
$\zeta_1=\frac{1}{\ln(2)}\sum\limits_{k=1}^{K}\sum\limits_{i=1}^{K}\sum\limits_{t=1}^{\omega_{k,i}}\Xi_K\Bigg(i,t,\{\omega_{k,q}\}_{q=1}^{K},\left\{\frac{\varrho_{k,q}}{\omega_{k,i}}\right\}_{q=1}^{K},\\\{l_{k,q}\}_{q=1}^{K-2}\Bigg)\ln\left(P2^{-\frac{B_i}{N_tN_r-1}}\alpha_{k,i}\right)$
and
$\zeta_2=\frac{1}{\ln(2)}\sum\limits_{k=1}^{K}\sum\limits_{i=1}^{K}\sum\limits_{t=1}^{\omega_{k,i}}\Xi_K\left(i,t,\{\omega_{k,q}\}_{q=1}^{K},\left\{\frac{\varrho_{k,q}}{\omega_{k,i}}\right\}_{q=1}^{K},\{l_{k,q}\}_{q=1}^{K-2}\right)$.
Since
$\frac{\partial\Delta\bar{R}(\textbf{B}_k,\textbf{d})}{\partial
d}>0$, $\Delta\bar{R}(\textbf{B}_k,\textbf{d})$ is an increasing
function of $d$, we have the following theorem:

\emph{Theorem 3}: In the case of large $P$ and $B$, $d=1$ is the
asymptotically optimal transmission mode.

In fact, it is easy to understand that when the interference is so
strong, single data stream transmission can decrease the residual
interference significantly, and thus improves the performance.

\subsection{Noise Limited Case}
If the interference term is negligible with respect to the noise
term due to the low transmit power, then the SINR is reduced as
\begin{equation}
\gamma_{k,j}=\frac{\kappa_{k,k}\left|\textbf{h}_{k,k}^H\textbf{T}_{j,j}^{(k,k)}\right|^2}{\sigma^2},\label{eqn43}
\end{equation}
which is equivalent to the interference-free case. As discussed
earlier,
$\kappa_{k,k}\left|\textbf{h}_{k,k}^H\textbf{T}_{j,j}^{(k,k)}\right|^2$
is $\kappa_{k,k}\chi^2(2)$ distributed, then the average
transmission rate can be computed as
\begin{eqnarray}
\bar{R}_{k,j}&=&\int_0^{\infty}\log_2\left(1+\frac{x}{\sigma^2}\right)\frac{\exp(-x/\kappa_{k,k})}{\kappa_{k,k}}dx\nonumber\\
&=&-\exp\left(\frac{\sigma^2}{\kappa_{k,k}}\right)\textmd{E}_{\textmd{i}}\left(-\frac{\sigma^2}{\kappa_{k,k}}\right).\label{eqn44}
\end{eqnarray}
It is found that $\bar{R}_{k,j}$ is independent of $B_{k,i}$ for all
$i$. Thus, in the noise limited case, there is no need for CSI
feedback, as CSI is useless for performance improvement.

In fact, the noise limited case can be considered as
interference-free, so the interference CSI is immaterial. It has
also been shown in \cite{OIA} that at low SNR region, IA does not
perform well as compared to the other interference mitigation
schemes. It is easy to derive the sum of average transmission rate
based on (\ref{eqn44}) as follows
\begin{eqnarray}
\bar{R}=\sum\limits_{k=1}^K-d_k\exp\left(\frac{\sigma^2}{\kappa_{k,k}}\right)\textmd{E}_{\textmd{i}}\left(-\frac{\sigma^2}{\kappa_{k,k}}\right).\label{eqn45}
\end{eqnarray}
Note that (\ref{eqn45}) is an increasing function of $d_k$, so we
also have the following theorem:

\emph{Theorem 4}: It is optimal to use the maximum $d_k$ fulfilling
the feasibility conditions of IA in the noise limited case.

As a simple example, for a symmetric MIMO interference network,
$d=\left\lfloor\frac{N_t+N_r}{K+1}\right\rfloor$ is optimal under
noise limited scenario. It is also aligned with the intuition that
in an interference-free network, the spatial multiplexing gain
should be exploited as much as possible. Similar phenomenon has also
been observed in traditional multiuser downlink networks \cite{MS}
\cite{Modeselection}.

\section{Numerical Results}
To evaluate the accuracy of the performance analysis results, and
the effectiveness of the performance optimization schemes for IA
under a limited feedback MIMO interference networks, we present
several simulation results under several different scenarios. For
convenience, we set $N_t=8$, $N_r=8$, $K=4$, $B=20$, $\sigma^2=1$,
$d_1=d_2=d_3=d_4=d=2$ and $\alpha_{i,j}$ given in Tab.\ref{Tab1} for
all simulation scenarios without explicit explanation. In addition,
we use SNR (in dB) to represent $10\log_{10}\frac{P}{\sigma^2}$.
Without loss of generality, we take the sum of average transmission
rate as the performance metric. We compare the proposed adaptive
feedback allocation scheme (PAS) with two baseline schemes, namely
equal feedback allocation scheme (EAS) and residual interference
minimization based feedback allocation scheme (RIMS). As the name
implies, EAS lets $B_{k,i}=B/K$ for all $k$ and $i$, and RIMS
distributes the feedback bits based on the criterion of the
minimization of average residual interference. Moreover, we also
compare the performance of dynamic mode selection scheme and fixed
mode scheme.

\newcommand{\tabincell}[2]{\begin{tabular}{@{}#1@{}}#2\end{tabular}}
\begin{table}\centering
\caption{Parameter Table for $\alpha_{i,j}, \forall i,j\in[1,4]$}
\label{Tab1}
\begin{tabular*}{4.92cm}{|c|c|c|c|c|}\hline
\backslashbox{i}{j}& 1 & 2 & 3& 4\\
\hline 1 &  1.00  & 0.50 & 0.10 & 0.01 \\
\hline 2 & 0.55 & 1.00 & 0.45 & 0.10 \\
\hline 3 & 0.10 & 0.45 & 1.00 & 0.55 \\
\hline 4 & 0.01 & 0.10 & 0.50 & 1.00 \\
\hline
\end{tabular*}
\end{table}


\begin{figure}[h] \centering
\includegraphics [width=0.5\textwidth] {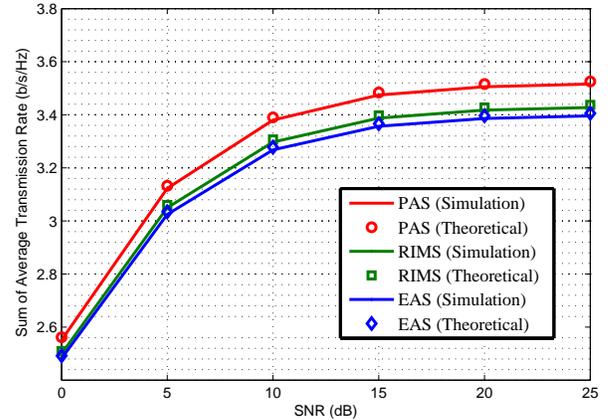}
\caption {Performance comparison of different feedback bits
allocation schemes} \label{Fig2}
\end{figure}

First, we compare the sum of average transmission rates of PAS,
RIMS, and EAS. Note that we present both theoretical and simulation
results for all the three schemes. As seen in Fig.{\ref{Fig2}}, the
theoretical results nearly coincide with the simulation results in
the whole SNR region, which testifies the high accuracy. From the
performance point of view, PAS performs better than RIMS and EAS,
since RIMS only considers the residual interference and EAS
completely ignores the channel conditions. As SNR increases, the
performance gain enlarges gradually. Therefore, PAS is an effective
performance optimization scheme for IA with limited CSI feedback in
the sense of maximizing the sum rate. It is worth pointing out that,
with respect to PAS, EAS and RIMS have lower complexity.
Specifically, EAS distributes the feedback bits equally, so the
computational complexity is $\mathcal{O}(1)$. RIMS computes the
feedback bits for each transmitter by maximizing the average
residual interference \cite{FeedbackAllocation1}
\cite{FeedbackAllocation2}, thus the computational complexity is
$\mathcal{O}(K)$. PAS allocates each feedback bit by comparing $K$
rate increments, so the computational complexity is
$\mathcal{O}(KB)$. Clearly, the performance gain is achieved at the
cost of complexity. In additional, it is found that there exist
performance ceilings for all the three schemes in the high SNR
region, which reconfirms Theorem 1. Furthermore, when the SNR is
low, all three schemes have nearly the same performance, since the
CSI feedback is useless in noise limited case as analyzed.

\begin{figure}[h] \centering
\includegraphics [width=0.5\textwidth] {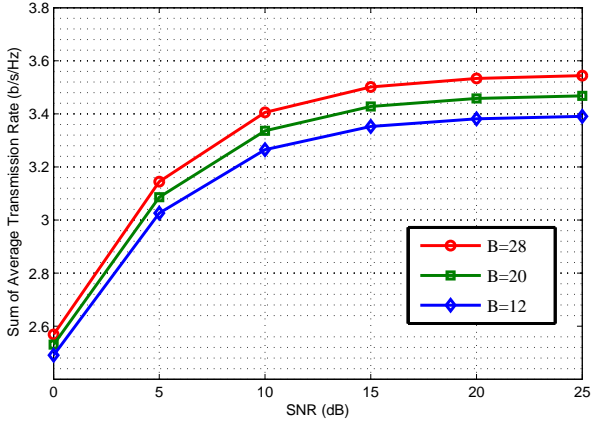}
\caption {Performance comparison of PAS with different feedback
bits} \label{Fig3}
\end{figure}

Secondly, we show the benefit of PAS from the perspective of
feedback amount. As seen in Fig. \ref{Fig3}, the performance gain
from higher feedback amount becomes larger with the increase of SNR.
Moreover, it is found that there always exists a performance ceiling
for a given $B$ after a saturation point, but the ceiling will rise
as $B$ increases. Thus, in order to avoid the ceiling, one should
increase $B$ according to the claim in Theorem 2. In addition, there
is hardly any performance gain in low SNR region even one increases
$B$, which reconfirms the claim that CSI is useless at low SNR.

Next, we investigate the impact of the number of transmit antenna
$N_t$ on the sum of average transmission rate for PAS. For a given
feedback amount $B$, the performance degrades with the increase of
the number of transmit antennas, this is because the quantization
accuracy decreases, and the residual interference increases
accordingly as explained in Theorem 2.

\begin{figure}[h] \centering
\includegraphics [width=0.5\textwidth] {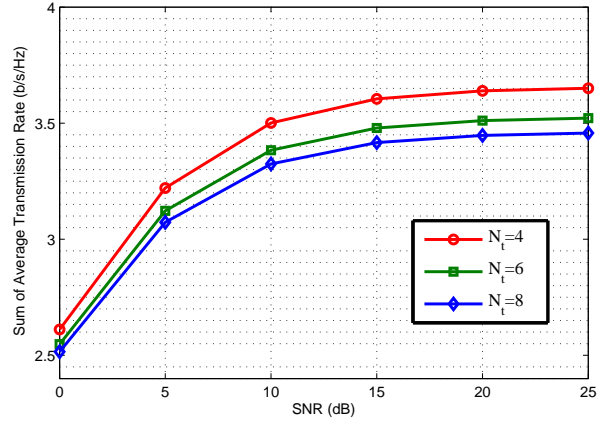}
\caption {Performance comparison of LBS scheme with different number
of transmit antennas and $B$=21} \label{Fig4}
\end{figure}

\begin{figure}[h] \centering
\includegraphics [width=0.5\textwidth] {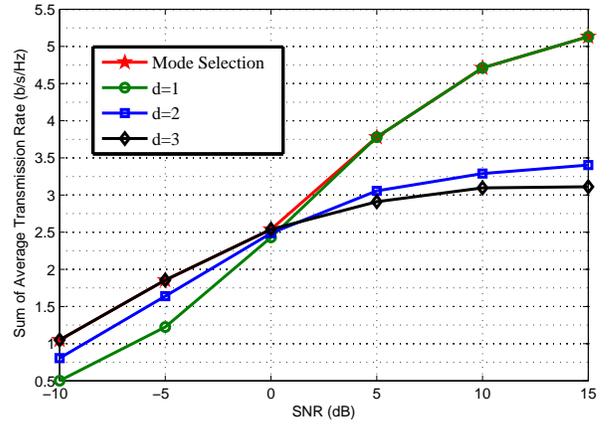}
\caption {Performance comparison of mode selection scheme with fixed
mode selection} \label{Fig5}
\end{figure}

Then, we compare the performance of dynamic mode selection scheme
and fixed mode scheme based on PAS. As shown in Fig.\ref{Fig5},
dynamic mode selection scheme always obtains the optimal
performance. For example, at SNR=10dB, dynamic mode selection scheme
can get about $1.5$b/s/Hz gain with respect to fixed mode scheme of
$d=2$. Thus, dynamic mode selection is a powerful performance
optimization for IA with limited CSI feedback. Moreover, it is found
that in the low SNR region, maximum feasible $d$ is chosen; while in
the high SNR region, $d=1$ is optimal. These validate the claims in
Theorem 4 and Theorem 3, respectively. More importantly, it is
illustrated that only two transmission modes are adopted possibly,
so we only need to compares the performances of the two modes when
performing mode selection, which reduces the complexity
significantly.

\begin{figure}[h] \centering
\includegraphics [width=0.5\textwidth] {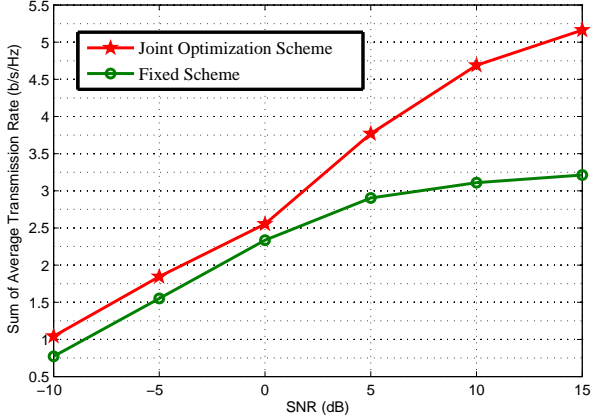}
\caption {Performance comparison of combined optimization scheme
with fixed scheme} \label{Fig6}
\end{figure}

Finally, we show the benefit of the proposed feedback allocation and
mode selection joint optimization scheme over the EAS with $d=2$. As
seen from Fig.\ref{Fig6}, even at very low SNR, the proposed
combined optimization scheme can achieve significant performance
gain. This is because the joint optimization scheme chooses the
maximum transmission mode, which is optimal under this condition.
With the increase of SNR, the performance gain becomes larger. For
example, there is an about 2b/s/Hz gain at SNR=15dB. The performance
gain comes from two folds: at high SNR, the network is interference
limited, the joint optimization scheme selects single stream mode to
decrease the interference; on the other hand, based on the optimal
single stream mode, the joint optimization scheme uses all feedback
bits to mitigate the inter-link interference, which increases the
feedback utility efficiency. Hence, the proposed joint optimization
scheme can effectively improve the performance of IA with limited
CSI feedback.

\section{Conclusion}
This paper addresses the problem of performance analysis and
optimization for IA in a general MIMO interference network with
limited CSI feedback. A major contribution of this paper is having
derived the closed-form expression of the average transmission rate
in terms of feedback amount and transmission mode. Based on this
result, we propose two feasible and effective performance
optimization schemes, namely adaptive feedback allocation and
dynamic mode selection. In addition, asymptotic analysis is carried
out to obtain further insights on system performance and design
guidelines. For example, our asymptotic results show that the number
of feedback bits has to be increased when the transmit power is
increased or when the number of antennas is increased; while under
noise-limited scenario, the feedback of CSI have no impact on
improving the system performance with IA, hence spatial degree of
freedom should be exploited as much as possible. On the contrary,
single data stream should be chosen in interference-limited
scenario.

\begin{appendices}
\section{The derivation of $\bar{R}_{k,j}(\textbf{B}_k,\mathbf{d})$}
Let $W(\textbf{B}_k,\textbf{d})$ and $V(\textbf{B}_k,\textbf{d})$
denote the first and second terms at the right hand of
(\ref{eqn25}). At first, we put the focus on
$W(\textbf{B}_k,\textbf{d})$. By substituting (\ref{eqn9}) into
(\ref{eqn25}), we have
\begin{eqnarray}
W&&\!\!\!\!\!\!\!\!\!\!\!\!\!\!\!(\textbf{B}_k,\textbf{d})\nonumber\\
&=&\frac{1}{\ln2}E\bigg[\ln\bigg(\kappa_{k,k}\left|\textbf{h}_{k,k}^H\textbf{T}_{j,j}^{(k,k)}\right|^2\nonumber\\
&&+\kappa_{k,k}a_{k,k}\|\textbf{h}_{k,k}\|^2\sum\limits_{l=1,l\neq
j}^{d_k}|\textbf{s}_{k,k}^H\textbf{T}_{j,l}^{(k,k)}|^2\nonumber\\
&&+\sum\limits_{i=1, i\neq
k}^{K}\kappa_{k,i}a_{k,i}\sum\limits_{l=1}^{d_i}\|\textbf{h}_{k,i}\|^2
\left|\textbf{s}_{k,i}^H\textbf{T}_{j,l}^{(k,i)}\right|^2+\sigma^2\bigg)\bigg].\nonumber\\\label{eqn29}
\end{eqnarray}
According to the theory of quantization cell approximation
\cite{RVQ}, $a_{k,i}\|\textbf{h}_{k,i}\|^2$ is
$\Gamma(N_tN_r-1,2^{-\frac{B_{k,i}}{N_tN_r-1}})$ distributed.
Moreover, $|\textbf{s}_{k,i}^H\textbf{T}_{j,l}^{(k,i)}|^2$ for
$i=1,\cdots,K$ are i.i.d. $\beta(1,N_tN_r-2)$ distributed, since
$\textbf{T}_{j,l}^{(k,i)}$ of unit norm is independent of
$\textbf{s}_{k,i}$. For the product of a
$\Gamma(N_tN_r-1,2^{-\frac{B_{k,i}}{N_tN_r-1}})$ distributed random
variable and a $\beta(1,N_tN_r-2)$ distributed random variable, it
is equal to $2^{-\frac{B_{k,i}}{N_tN_r-1}}\chi^2(2)$ in distribution
\cite{QCA}. Based on the fact that the sum of $M$ i.i.d. $\chi^2(2)$
distributed random variables is $\chi^2(2M)$ distributed,
$a_{k,i}\sum\limits_{l=1}^{d_i}\|\textbf{h}_{k,i}\|^2
\left|\textbf{s}_{k,i}^H\textbf{T}_{j,l}^{(k,i)}\right|^2$ is
$2^{-\frac{B_{k,i}}{N_tN_r-1}}\chi^2(2d_i)$ distributed.
Additionally, since $\textbf{T}_{j,j}^{(k,k)}$ is designed
independently of $\textbf{h}_{k,k}$,
$\left|\textbf{h}_{k,k}^H\textbf{T}_{j,j}^{(k,k)}\right|^2$ is also
$\chi^2(2)$ distributed. Hence,
$\kappa_{k,k}\left|\textbf{h}_{k,k}^H\textbf{T}_{j,j}^{(k,k)}\right|^2+\kappa_{k,k}a_{k,k}\|\textbf{h}_{k,k}\|^2\sum\limits_{l=1,l\neq
j}^{d_k}|\textbf{s}_{k,k}^H\textbf{T}_{j,l}^{(k,k)}|^2+\sum\limits_{i=1,
i\neq
k}^{K}\kappa_{k,i}a_{k,i}\sum\limits_{l=1}^{d_i}\|\textbf{h}_{k,i}\|^2
\left|\textbf{s}_{k,i}^H\textbf{T}_{j,l}^{(k,i)}\right|^2$ is a
nested finite weighted sum of $K+1$ Erlang pdfs, whose pdf is given
by \cite{PDF}
\begin{eqnarray}
f(x)&=&\sum\limits_{i=1}^{L}\sum\limits_{t=1}^{\eta_{k,i}}\Xi_{L}\left(i,t,\{\eta_{k,q}\}_{q=1}^{L},\left\{\frac{\varrho_{k,q}}{\eta_{k,q}}\right\}_{q=1}^{L},\{l_{k,q}\}_{q=1}^{L-2}\right)\nonumber\\
&&\times
g\left(x,t,\frac{\varrho_{k,i}}{\eta_{k,i}}\right),\label{eqn30}
\end{eqnarray}
where $L=K+1$, $\eta_{k,i}=d_i$,
$\varrho_{k,i}=\kappa_{k,i}2^{-\frac{B_{k,i}}{N_tN_r-1}}$ for all
$i\neq k$, $\eta_{k,k}=d_k-1$,
$\varrho_{k,k}=\kappa_{k,k}2^{-\frac{B_{k,k}}{N_tN_r-1}}$,
$\eta_{k,L}=1$, $\varrho_{k,L}=\kappa_{k,k}$, and
$g(x,t,\varrho_{k,i})=\frac{x^{t-1}\exp(-x/\varrho_{k,i})}{\varrho_{k,i}^{t}\Gamma(t)}$
for all $i\in[1,L]$. The weights $\Xi_{L}$ are defined as
\begin{eqnarray}
\Xi_{L}\!\!\!\!\!\!\!&&\!\!\!\!\!\!\!\left(i,t,\{\eta_{k,q}\}_{q=1}^{L},\{\varrho_{k,q}\}_{q=1}^{L},\{l_{k,q}\}_{q=1}^{L-2}\right)\nonumber\\
&=&\sum\limits_{l_{k,1}=t}^{\eta_{k,i}}\sum\limits_{l_{k,2}=t}^{l_{k,1}}\cdots\sum\limits_{l_{k,L-2}=t}^{l_{k,L-3}}\bigg[\frac{(-1)^{T_L-\eta_{k,i}}\varrho_{k,i}^{t}}{\prod_{h=1}^{L}\varrho_{k,h}^{\eta_{k,h}}}\nonumber\\
&&\times\frac{\Gamma(\eta_{k,i}+\eta_{k,1+\textmd{U}(1-i)}-l_{k,1})}{\Gamma(\eta_{k,1+\textmd{U}(1-i)})\Gamma(\eta_{k,i}-l_{k,1}+1)}\nonumber\\
&&\times\left(\frac{1}{\varrho_{k,i}}-\frac{1}{\varrho_{k,1+\textmd{U}(1-i)}}\right)^{l_{k,1}-\eta_{k,i}-\eta_{k,1+\textmd{U}(1-i)}}\nonumber\\
&&\times\frac{\Gamma(l_{k,L-2}+\eta_{k,L-1+\textmd{U}(L-1-i)}-t)}{\Gamma(\eta_{k,L-1+\textmd{U}(L-1-i)})\Gamma(l_{L-2}-t+1)}\nonumber\\
&&\times\left(\frac{1}{\varrho_{k,i}}-\frac{1}{\varrho_{k,L-1+\textmd{U}(L-1-i)}}\right)^{t-l_{k,L-2}-\eta_{k,L-1+\textmd{U}(L-1-i)}}\nonumber\\
&&\times\prod_{s=1}^{L-3}\frac{\Gamma(l_{k,s}+\eta_{k,s+1+\textmd{U}(s+1-i)}-l_{k,s+1})}{\Gamma(\eta_{k,s+1+\textmd{U}(s+1-i)})\Gamma(l_{k,s}-l_{k,s+1}+1)}\nonumber\\
&&\times\left(\frac{1}{\varrho_{k,i}}-\frac{1}{\varrho_{k,s+1+\textmd{U}(s+1-i)}}\right)^{l_{k,s+1}-l_{k,s}-\eta_{k,s+1+\textmd{U}(s+1-i)}}\bigg],\nonumber\\\label{eqn31}
\end{eqnarray}
where $T_L=\sum_{i=1}^{L}\eta_{k,i}$ and $\textmd{U}(x)$ is the
well-known unit step function defined as $\textmd{U}(x\geq0)=1$ and
zero otherwise. Note that the weights $\Xi_L$ are constant when
given $\eta_{k,i}$ and $\varrho_{k,i}$. Thus,
$W(\textbf{B}_k,\textbf{d})$ can be computed as
\begin{eqnarray}
W&&\!\!\!\!\!\!\!\!\!\!\!\!\!\!\!(\textbf{B}_k,\textbf{d})\nonumber\\&=&\frac{1}{\ln2}\int_{0}^{\infty}\ln(x+\sigma^2)f(x)dx\nonumber\\
&=&\frac{1}{\ln2}\sum\limits_{i=1}^{L}\sum\limits_{t=1}^{\eta_{k,i}}\Xi_L\Bigg(i,t,\{\eta_{k,q}\}_{q=1}^{L},\left\{\frac{\varrho_{k,q}}{\eta_{k,q}}\right\}_{q=1}^{L},\nonumber\\
&&\{l_{k,q}\}_{q=1}^{L-2}\Bigg)\int_{0}^{\infty}\ln(x+\sigma^2)g\left(x,t,\frac{\varrho_{k,i}}{\eta_{k,i}}\right)dx\nonumber\\
&=&\frac{1}{\ln2}\sum\limits_{i=1}^{L}\sum\limits_{t=1}^{\eta_{k,i}}\Xi_L\Bigg(i,t,\{\eta_{k,q}\}_{q=1}^{L},\left\{\frac{\varrho_{k,q}}{\eta_{k,q}}\right\}_{q=1}^{L},\nonumber\\
&&\{l_{k,q}\}_{q=1}^{L-2}\Bigg)Z\left(\sigma^2,t,\frac{\varrho_{k,i}}{\eta_{k,i}}\right),\label{eqn32}
\end{eqnarray}
where
\begin{eqnarray}
Z&&\!\!\!\!\!\!\!\!\!\!\!\!\!\!\!\left(\sigma^2,t,\frac{\varrho_{k,i}}{\eta_{k,i}}\right)\nonumber\\&=&\int_{0}^{\infty}\ln(x+\sigma^2)g\left(x,t,\frac{\varrho_{k,i}}{\eta_{k,i}}\right)dx\nonumber\\
&=&\ln(\sigma^2)+\sum\limits_{\vartheta=0}^{t-1}\frac{1}{\Gamma(t-\vartheta)}\bigg((-1)^{t-\vartheta-2}\left(\frac{\sigma^2\eta_{k,i}}{\varrho_{k,i}}\right)^{t-\vartheta-1}
\nonumber\\&&\times\exp\left(\frac{\sigma^2\eta_{k,i}}{\varrho_{k,i}}\right)\textmd{E}_{\textmd{i}}\left(-\frac{\sigma^2\eta_{k,i}}{\varrho_{k,i}}\right)\nonumber\\
&&+\sum\limits_{\nu=1}^{t-\vartheta-1}\Gamma(\nu)\left(-\frac{\sigma^2\eta_{k,i}}{\varrho_{k,i}}\right)^{t-\vartheta-\nu-1}\bigg),\label{eqn53}
\end{eqnarray}
and
$\textmd{E}_{\textmd{i}}(x)=\int_{-\infty}^{x}\frac{\exp(t)}{t}dt$
is the exponential integral function. (\ref{eqn53}) follows from
[30, Eq. 4.3375].

Note that $V(\textbf{B}_k,\textbf{d})$ is similar to
$W(\textbf{B}_k,\textbf{d})$ except the absence of the term
$\kappa_{k,k}\left|\textbf{h}_{k,k}^H\textbf{T}_{j,j}^{(k,k)}\right|^2$.
Hence, $V(\textbf{B}_k,\textbf{d})$ can be expressed as
\begin{eqnarray}
V(\textbf{B}_k,\textbf{d})
&=&\frac{1}{\ln2}\sum\limits_{i=1}^{K}\sum\limits_{t=1}^{\omega_{k,i}}\Xi_K\Bigg(i,t,\{\omega_{k,q}\}_{q=1}^{K},\left\{\frac{\varrho_{k,q}}{\omega_{k,i}}\right\}_{q=1}^{K},\nonumber\\
&&\{l_{k,q}\}_{q=1}^{K-2}\Bigg)Z\left(\sigma^2,t,\frac{\varrho_{k,i}}{\omega_{k,i}}\right),\label{eqn33}
\end{eqnarray}
where $\omega_{k,k}=d_k-1$, $\omega_{k,i}=d_i$ for $i\neq k$, and
$\varrho_{k,i}=\kappa_{k,i}2^{-\frac{B_{k,i}}{N_tN_r-1}}$ for all
$i\in[1,K]$. So, we obtain the average transmission rate for the
$j$th data stream of link $k$ as follows
\begin{eqnarray}
\bar{R}_{k,j}&&\!\!\!\!\!\!\!\!\!\!\!\!\!\!\!(\textbf{B}_k,\textbf{d})\nonumber\\
&=&W(\textbf{B}_k,\textbf{d})-V(\textbf{B}_k,\textbf{d})\nonumber\\
&=&\frac{1}{\ln2}\sum\limits_{i=1}^{L}\sum\limits_{t=1}^{\eta_{k,i}}\Xi_L\Bigg(i,t,\{\eta_{k,q}\}_{q=1}^{L},\left\{\frac{\varrho_{k,q}}{\eta_{k,i}}\right\}_{q=1}^{L},\nonumber\\
&&\{l_{k,q}\}_{q=1}^{L-2}\Bigg)Z\left(\sigma^2,t,\frac{\varrho_{k,i}}{\eta_{k,i}}\right)\nonumber
\end{eqnarray}
\begin{eqnarray}
&&-\frac{1}{\ln2}\sum\limits_{i=1}^{K}\sum\limits_{t=1}^{\omega_{k,i}}\Xi_K\Bigg(i,t,\{\omega_{k,q}\}_{q=1}^{K},\left\{\frac{\varrho_{k,q}}{\omega_{k,i}}\right\}_{q=1}^{K},\nonumber\\
&&\{l_{k,q}\}_{q=1}^{K-2}\Bigg)Z\left(\sigma^2,t,\frac{\varrho_{k,i}}{\omega_{k,i}}\right).\label{eqn46}
\end{eqnarray}

\section{The derivation of $\bar{R}_{k,j}(\textbf{B}_k,\mathbf{d})$ in interference limited case}
For convenience, we use $G(\textbf{B}_k,\textbf{d})$ and
$H(\textbf{B}_k,\textbf{d})$ to denote the first and second terms at
the right hand of (\ref{eqn38}). The pdf of
$\kappa_{k,k}\left|\textbf{h}_{k,k}^H\textbf{T}_{j,j}^{(k,k)}\right|^2+I_{k,j}$
is given by $f(x)$ in (\ref{eqn30}), thus
$G(\textbf{B}_k,\textbf{d})$ can be computed as
\begin{eqnarray}
G&&\!\!\!\!\!\!\!\!\!\!\!\!\!\!\!(\textbf{B}_k,\textbf{d})\nonumber\\
&=&\frac{1}{\ln2}\int_{0}^{\infty}\ln(x)f(x)dx\nonumber\\
&=&\frac{1}{\ln2}\sum\limits_{i=1}^{L}\sum\limits_{t=1}^{\eta_{k,i}}\Xi_L\Bigg(i,t,\{\eta_{k,q}\}_{q=1}^{L},\left\{\frac{\varrho_{k,q}}{\eta_{k,i}}\right\}_{q=1}^{L},\nonumber\\
&&\{l_{k,q}\}_{q=1}^{L-2}\Bigg)\int_{0}^{\infty}\ln(x)\frac{x^{t-1}\exp(-\frac{x}{\varrho_{k,i}/\eta_{k,i}})}{(\varrho_{k,i}/\eta_{k,i})^{t}\Gamma(t)}dx\nonumber\\
&=&\frac{1}{\ln2}\sum\limits_{i=1}^{L}\sum\limits_{t=1}^{\eta_{k,i}}\Xi_L\Bigg(i,t,\{\eta_{k,q}\}_{q=1}^{L},\left\{\frac{\varrho_{k,q}}{\eta_{k,i}}\right\}_{q=1}^{L},\nonumber\\
&&\{l_{k,q}\}_{q=1}^{L-2}\Bigg)\left(\psi(t)+\ln(\varrho_{k,i})-\ln(\eta_{k,i})\right),\label{eqn39}
\end{eqnarray}
where $\psi(x)=\frac{d\ln(\Gamma(x))}{dx}$ is the Euler psi
function. (\ref{eqn39}) is obtained based on [30, Eq. 4.352.1].
Similarly, we can get $H(\textbf{B}_k,\textbf{d})$ as follows:
\begin{eqnarray}
H(\textbf{B}_k,\textbf{d})&=&\frac{1}{\ln2}\sum\limits_{i=1}^{K}\sum\limits_{t=1}^{\omega_{k,i}}\Xi_K\Bigg(i,t,\{\omega_{k,q}\}_{q=1}^{K},\left\{\frac{\varrho_{k,q}}{\omega_{k,i}}\right\}_{q=1}^{K},\nonumber\\
&&\{l_{k,q}\}_{q=1}^{K-2}\Bigg)\left(\psi(t)+\ln(\varrho_{k,i})-\ln(\omega_{k,i})\right).\label{eqn40}
\end{eqnarray}
Hence, the average transmission rates for the $j$th data stream of
link $k$ is given by
\begin{eqnarray}
\bar{R}_{k,j}&&\!\!\!\!\!\!\!\!\!\!\!\!\!\!\!(\textbf{B}_k,\textbf{d})\nonumber\\
&=&\frac{1}{\ln2}\bigg(\sum\limits_{i=1}^{L}\sum\limits_{t=1}^{\eta_{k,i}}\Xi_L\Bigg(i,t,\{\eta_{k,q}\}_{q=1}^{L},\left\{\frac{\varrho_{k,q}}{\eta_{k,i}}\right\}_{q=1}^{L-2},\nonumber\\
&&\{l_{k,q}\}_{q=1}^{L-2}\Bigg)\left(\psi(t)+\ln(\varrho_{k,i})-\ln(\eta_{k,i})\right)\nonumber\\
&-&\sum\limits_{i=1}^{K}\sum\limits_{t=1}^{\omega_{k,i}}\Xi_K\Bigg(i,t,\{\omega_{k,q}\}_{q=1}^{K},\left\{\frac{\varrho_{k,q}}{\omega_{k,i}}\right\}_{q=1}^{K},\nonumber\\
&&\{l_{k,q}\}_{q=1}^{K-2}\Bigg)\left(\psi(t)+\ln(\varrho_{k,i})-\ln(\omega_{k,i})\right)\bigg),\label{eqn47}
\end{eqnarray}
\end{appendices}

%
%

\end{document}